# From Algorithm Worship to the Art of Human Learning: Insights from 50-year journey of AI in Education

Opinion

Kaśka Porayska-Pomsta

University College London, UCL Knowledge Lab

## 1. Introduction

Over the past decade, there have been increasing proclamations from diverse stakeholders that humanity is at an inflection point due to advances in Artificial Intelligence (AI) technologies (e.g., Crawford, 2017). The general public are conditioned by this messaging to expect both big (though so far largely non-descript) changes to our lives, including to the way that we learn and teach. We are conditioned to accept the inevitability of AI in our personal and professional activities[1]. Hopes have been highlighted for how AI might change our lives for the better, for example, by helping us unlock some of the greatest scientific mysteries needed for advancing healthcare (Jumper et al., 2021). Warnings have been also articulated regarding whether and how AI might fundamentally change the way we perceive reality, how we form our beliefs, or interact with one another (Bostrom, 2017). More recently, questions started to emerge about AI's transformative potential (for better or worse) for our functioning at neurocognitive, socio-emotional, individual and collective levels (UNESCO, 2022; Pedro, et al., 2019, Porayska-Pomsta, 2023), along with concerns regarding the ethical implications of using AI for supporting human decision-making in contexts that are both high-stakes (e.g., for medical diagnoses or for student assessment) and relatively low-stakes, e.g., selecting movies on streaming sites.

Such *hope-fear* rhetoric is also present in the context of AI applications to supporting human learning in formal and informal contexts. Recent hopes for AI in education (AIED) largely relate to delivering learning at scale across different geographical and cultural contexts, especially in light of growing global teacher shortages and diminishing funding for education in many countries (UNESCO, 2023). These hopes are increasingly used to fuel politically and market motivated discourse about the need to 'release teachers from tedious tasks' such as standardised assessments to allow them to focus on the 'things that matter' (Gentile et al., 2023), or to justify the narrowing of the formal education curricula mainly to STEM subjects. Such subjects tend to be perceived as having more immediately quantifiable outcomes and to be more useful economically than, for example, arts and humanities. The *efficiency*[2] and *measurability* rhetoric around Education and its purpose is deeply ingrained in both the economics of the education sector and the emergent economics of AI (e.g., Mazzucato, 2022), and as such, it seems to dominate the current mainstream discussions about AIED and the future of Education supported by AI. In this context, the questions around how technology might and should shape both what and how we learn and, by extension, what and how we teach, are increasingly prominent in the emergent debates and recommendations at national and international levels[3] (Jobin,

---

[1] https://medium.com/@sumsonline/welcoming-the-inevitable-impact-of-ai-on-education-98b4697e4a35

[2] Note that efficiency, which means doing something with the minimum waste of time and effort, and which emphasises benefits for the Education system as reflecting specific economic and politic interests, is often conflated with efficacy and effectiveness, which refer respectively to: (i) the ability of something like an education intervention to produce desired results and (ii) an intervention actually producing desired results, and which put the emphasis on the benefits for the learners.

[3] https://www.unesco.org/en/digital-education/artificial-intelligence

Ienca, & Vayena, 2019). Unfortunately, the way that these questions are formulated remains at such a high level of generality that it is hard to contextualise or address them in meaningful ways that can be backed with corresponding individual or collective actions.

In this article, I briefly examine the key reasons for why the discussions about the role and the potential of AI in Education remain vague and I outline some of the most concerning consequences of this vagueness for the way that we formulate and address questions about the future of Education in view of AI. I then highlight insights from educational research of pertinence to our understanding of the first principles of learning and teaching. This will be followed by a brief synthesis of the approaches to AIED which explore the application of AI for implementing these principles. The paper concludes with an overarching discussion of questions that are important to shaping our thinking and action in the context of AI-supported learning and teaching.

## 2. Going round in circles: reasons for why we currently find it difficult to make the most of AIED

There are several intertwined reasons for why we experience significant challenges in formulating precisely the questions concerning Education's purpose, values, and the overarching Education system in the face of AI as a potential game-changer for the sector.

First, beyond the field itself, there's a marked ignorance and neglect of AI in Education (AIED) as a mature domain of research and practice. Much of this research and practice falls under the International Society for Artificial Intelligence in Education's purview (https://iaied.org/). Even though the AIED research community has been producing in-depth knowledge on AI engineering for human learning, human-computer interaction for education, and AIED efficacy evaluations for nearly 50 years, these insights are rarely known or considered in broader discussions about Education's future involving AI (see e.g., the keynotes at the last UNESCO Digital Learning Week[4]). This omission indicates a worrying disconnect between the critical knowledge from AIED research and concrete AI technologies, their implementations at the front-line of human learning, and the broader AI research, the EdTech industry, and policy.

Second, there's a prevalent inclination in multi-stakeholder discussion to perceive AI as a uniform artefact and to assume its utility and impact to be consistent across varied contexts and user groups. This inclination ignores the fact that while at a theoretical level, AI can be uniformly defined as autonomously pursuing predefined objectives and as able to learn from data, specific AI systems exhibit diverse characteristics. These systems differ in their design, purpose, operational data, inference mechanisms, target users, and ultimate objectives. For instance, AI tools in the medical field, despite often employing similar techniques as those in education, differ vastly in their purpose, usage, and success metrics. Even within the AIED domain, AI systems show variance based on target users, contexts of usage, learning objectives, pedagogical foundations, evaluation criteria and implementation specifics.

The third issue relates to a widespread oversimplification of the 'Education' construct outside the broad domains of the Learning Sciences and AIED research. Many mistakenly perceive Education as a catch-all term for any form of learning, regardless of the individual learners or socio-cultural settings. However, Education is multifaceted, influenced by myriad administrative, socio-cultural, and policy dimensions, value systems, curriculum designs, pedagogies, localised contexts, diverse learners' experiences and needs, and educators with varied skills and motivations. Additionally, the science of human learning encompasses a large number of different dimensions from the human development and psychology, subject domains ontologies, pedagogical approaches, communication, to philosophy and sociology of education.  Thus, generalising 'Education' as a monolithic concept and on that basis attempting to forecast the trajectory of AI's role and nature in Education as a uniform entity is akin to relying on a daily horoscope for life-altering decisions. In this context a prevalent misconception is that everyone, having been part of an educational process at some point, inherently *understands* what 'Education' is and how it works. Yet, individual experiences with Education differ drastically based on personal, socio-cultural, and national contexts. Contrary to the widespread notion that everyone understands what 'Education' is, many, including notable AI experts[5], unfortunately lack a nuanced understanding required for constructive discussions on AI-enhanced future of Education and indeed, since they demonstrably fail to engage with the AIED and the Learning Sciences research and practice, they also lack the knowledge about what is already and will be in the future genuine innovation in this context.

---

[4] https://www.unesco.org/en/weeks/digital-learning

Forth is the issue of differing incentives of relevance to the diverse 'Education' stakeholders (Nti Asare, 2023), which is driven by the stakeholders' multiple and often conflicting interests and expectations (Dubnik, 2014).  This issue ultimately hinders our collective ability to determine clearly and definitively the purpose of Education and by extension of AI in Education at a collectively actionable level.  While in general, we can talk about certain values as being encoded in what and how we teach, different stakeholders will be incentivised by very different things: e.g., politicians – by their ability to demonstrate their policies' success and to attract votes; EdTech entrepreneurs – by their ability to sell their EdTech products and keep their businesses afloat, or indeed, to make a profit; parents – by their desire to see their children succeed academically so that they progress through education and become successful independent adults; teachers and educators – by a diversity of motivations from their vocation to engender learning and help individuals become the best they can be, and/or to develop a respectable career for themselves, and learners – by the adult-defined prerogatives to simply go to school (if they are school-aged children), or by a necessity to gain qualifications to progress in their jobs or to further their special interests (if they are adults). Even this general overview of why different individuals may be incentivised to engage in 'Education' remains quite vague, as such continuing to make it hard for us to pin down exactly what and whose values we are to encode in our definitions of Education and, by extension – in our definition of AI's role in Education.

## 3. AI in Education is not a WUMPUS problem: the consequences of ignorance and over-generalisation

The prevailing narrative that leans heavily on computer scientists who lack the specialised understanding that educational and AI-in-Education (AIED) researchers bring to the table, skews the discussion towards the techno-centric perspective. Within this perspective, 'Education' is treated analogously to a textbook AI problem, reminiscent of the Wumpus World (Russell & Norvig, 2020). In this board game-like challenge, AI must navigate a cave, symbolised by a grid, seeking gold and evading threats like stinky pits or the Wumpus monster. The Wumpus World is an AI staple because it mimics the decision-making complexities that intelligent systems face in real-world scenarios, using limited or indirect data.

While AI methods used to tackle the Wumpus-like challenges have been routinely applied in AI for Education (Woolf, 2008), they have been also documented for their limitations in those contexts. Specifically, real-world AIED solutions need to be context-sensitive, to focus on and respond to individual and fine-grained user needs (including their likely cognitive and emotional states) and on the specific outcomes of user interaction.  They also incorporate nuanced and evidence-based understanding of domain-dependent interaction components, such as if, what, when and how to offer feedback to the learners to maximise both motivation and the learning itself, or how to assess the extent of learners' knowledge and motivation at any given point. These interaction components are necessary for guiding individual students through different knowledge domains and changes in their understanding of such domains, and for adapting the support offered. Unlike in the Wumpus challenge with its well-defined aims and success measures, the conclusions gleaned from interactions with AI systems in educational contexts may often lack the accuracy needed to support efficacious feedback and other learning support decisions, such as content selection. As such the AIED's community's in-depth appreciation of the complexities related to detecting the necessary data, interpreting these data for assessments, feedback and adaptation of a learning environment has led over the years to new ways of thinking about how AI and the interaction with it may be designed, deployed and evaluated for more efficacious learning (see section 5 for concrete examples). These new ways, which emphasise design flexibility and user control over the AI system, offer both educationally valuable approaches (theoretically grounded and evaluated empirically) and exciting new avenues for responsible AI more broadly (e.g., inherently relying for their educational value on the principles of transparency, explainability and user control). They also offer approaches which challenge in a constructive and actionable way the standard models of both AI and of Education (Porayska-Pomsta, 2023; more on the standard models in sections 4 and 5, respectively).

---

[5] https://www.theguardian.com/technology/2023/jul/07/ai-likely-to-spell-end-of-traditional-school-classroom-leading-expert-says

Whilst a bit of a caricature, the Wumpus example helps to underscore a concerning trend within the current discourse, which prioritises tech potential over genuine human needs. Given AI's inherent limitations, as noted by non-AIED experts like Russell (2019), our discussions and the corresponding conclusions, and actions risk being streamlined by mal-informed parties to the narrow, Wumpus-like view of AI in Education, rather than being rooted in authentic educational principles and societal values. As an analogue of the problem consider computer scientists telling health professionals how best to conduct their practices of diagnosing and treating patients, simply by virtue of the potential of the technologies they have built to enhancing medical practices. Such scenario would never pass within Medicine without proper oversight from relevant medical practitioners, extensive trials or evidence, public consultation, and involvement of appropriate ethics and regulatory boards. Yet such a scenario is currently being enacted routinely in Education without proper involvement of those who research and/or practice Education and AIED in its diverse forms.

Another consequence of the technocentric viewpoint exercised in most discussions about AIED to date, coupled with a reliance on individuals who lack the essential theoretical and hands-on knowledge of human learning and development, or of designing technology for such development, is that it shapes a distinct vision of what AI in Education technologies are. This vision leans heavily towards didactic, instructionist methods and reinforcement-based techniques for knowledge relay, which to non-AIED specialists often appears to be the 'low-hanging-fruit' of Education for their implementation through technology. Such methods, focused largely on drill-and-practice, have long been considered important, but insufficient to foster deep learning, ignite a lifelong passion to learn, or cater to a diversity of students in terms of their social, cultural, and neuro-developmental backgrounds.

In essence, this perspective promotes the idea of Education as a mere conduit for information, prioritising swift, measurable outcomes over the bold vision of endowing learners with the tools and independence to learn on their own terms. This skewed vision resonates with the model that values the efficiency of learning and teaching discussed earlier, at the expense of guided discovery and learners' active participation in the construction of their knowledge. Ironically, this approach contradicts the aspirational goals articulated by numerous policymakers who accept this view of AIED to arm learners with the so-called 21st-century skills like creativity, critical thinking, and foresight. The repercussions of this skewed emphasis are already apparent in the reports about young people's

(including school children's) declining well-being on multiple fronts, ranging from cognitive and emotional health, to socio-cultural and political dimensions (e.g., Clinton, 2023). Yet, despite often dismissive critique from non-education specialists, the concept of empowerment via learner-driven actions is not an esoteric ideal. The significance of such empowerment stands robustly validated across extensive educational research, with numerous examples, from both traditional educational practice and tangible AIED systems showcasing the feasibility and educational importance of this approach. Notably, there are many examples of AIED paradigms and their specific implementations that go against the grain of didactic approaches to supporting learning (see section 5 for a brief discussion of those) and that provide a window into how AI can efficaciously support the high-level cognitive and emotional competencies encapsulated in the 21-st century skills rhetoric *as well as* help us genuinely innovate the way we teach and learn (see also Porayska-Pomsta & Rajendran, 2019; Porayska-Pomsta, Holmes & Nemorin, 2023; Porayska-Pomsta, 2016).

A further implication is one which emerges from the previous two and it relates to *if* and *how* AIED is governed. The need for the governance of AI in general is a topic of active discussions at national and transnational levels. Numerous recommendations have been developed over the past few years, each providing ever increasing details about what aspects of AI may present particular risks and why. Multiple recommendations have been produced for the oversight and management of AI technologies, offering some universals of responsible and human-centred AI. UNESCO has produced a set of recommendations for AI in Education (UNESCO, 2021), and more recently for the use of generative AI use in Education more specifically (UNESCO, 2023b). Nevertheless, most recommendations lack the contextual, the theoretical or the practical detail of AIED to devise governing mechanisms needed for concrete action. Even if we can regulate for the universal ethical concerns regarding AI applications for education, without the relevant knowledge and insight we remain hindered in our ability to regulate AIED in a fully informed way that would cater the flexibility needed to respond to the changing technological landscape as it emerges. Simply, if we do not understand exactly what we are regulating for, we cannot regulate for it effectively and thus, remain back-footed into having to second-guess what will come next, and to react to (rather than plan for) any new developments.

A final, crucial concern that emerges as a direct consequence of our current limitations in creating governance rules for AIED is that without proper regulation, the field of AIED is open to misuse by those looking to make a profit from it. This means that Education, instead of being a public good, can be turned into a market where learning is monetised, where students and educators are treated as consumers, and where the products provided might be more about making money for EdTech entrepreneurs than about actually enhancing society's knowledge and critical skills. This should be highly alarming to everyone, given that these unregulated products can affect the mental and emotional development of learners (some as young as toddlers) in ways we might not fully understand or intend.

## 4. Lessons from Education

In order to progress in our understanding of the issues discussed thus far and in developing appropriate ways in which we can address them, it is essential to refer to the known principles of learning and teaching. In his 2018 keynote at the International Conference of AI in Education[6], Paulo Blinkstein recounted several of such principles as a way of highlighting the troubling prevalent trend within the broad EdTEch industry to oversimplify and commodify Education challenges for the purpose of promoting their products as 'solutions'. For example, these oversimplifications involve the misconception that teachers need rescuing from repetitive and boring tasks, such as assessment, so that they can focus more on actual teaching; that education is primarily about explanation; that learning is about continuous repetition; and that teaching is mainly about breaking content into tiny pieces (see also, Blinkstein, 2013).

---

[6] https://aied2018.utscic.edu.au/

Blinkstein contrasted these misconceptions with knowledge gleaned from decades of educational research. For example, he highlighted that what many consider boring tasks for teachers, such as administrative duties around assessment and reporting, actually form an integral part of the teachers' ability to develop in-depth understanding of each of their students' strengths and areas for improvement and as such, are foundational to their ability to offer tailored support. The fact that teachers are overworked is likely an outcome of under-funded educational systems in the face of increasingly over-crowded classrooms and diminishing incentives for potential new teachers to enter the profession and thus, decreasing teachers' numbers. In this context, the narrative promoting automation of certain 'routine' teaching tasks serves more the political agendas than the actual teachers and learners at the front-line.

In relation to the points regarding the nature of teaching and learning, Blinkstein points to a large body of multidisciplinary research that demonstrates unequivocally the crucial importance of learners' active participation in the construction of their knowledge. Here, explanation by teachers is important in guiding the direction of active exploration and enquiry by learners, but in and by itself it is insufficient to engender learning. Similarly, repetition of tasks and of information by learners provides a limited means for learning, which is fundamentally about constant contextualised equilibration between old and new theories held by individuals about the specific knowledge/learning domains.

Finally, in responding to the point about the low level of granularity of concepts needed by learners to 'understand' them, Blinkstein points to Papert's suggestions that complex concepts might be sometimes easier to learn than simpler ones, because when learners are engaged in meaningful, interest-driven, and hands-on activities, their intrinsic motivation and natural inclination for exploration will drive them to navigate through complex problem-solving and learning with ease and enthusiasm. Essentially, when learning is embedded in a context that is personal, explorative, and authentic, even complex ideas become accessible and exciting to learn, as they are intertwined with purpose and personal relevance.

Given a large body of educational research and well-documented front-line practices, learning is not a rigid thing that is confined to any specific environment such as classroom. Instead, it is a process that constantly evolves in tandem with the developing body and mind of an individual. As such, learning is environmentally, socio-culturally and psycho-neurologically conditioned, and it is also shaped by political and economic factors, as well as specific circumstances of individuals. Essentially, everything that we do, every little decision we make, every action we take, every interaction we have with the people and things around us *intervenes* with our perception of the world, and consequently, it transforms our thinking and behaviour, often in ways that are not observable immediately, but whose impact becomes apparent over time.

In formal education, which is guided by standardised curricula and assessments, pedagogy (in Papert's words – *the art of teaching*) has been perceived by many as the backbone of learning – a perception which has also guided the emphasis in the discussions around AI as an enhancer of teaching rather than of learning. Yet as research shows it is not teaching *per se* that engenders learning, but active construction of knowledge by learners themselves, i.e., *the art of learning* (Papert, 1994). Pedagogy, including appropriately selected and timed feedback, whilst being important, acts as enabler of the learning environment in which learning takes place. Within this perspective, i.e., one which elevates the art of learning to at least being as important as the art of teaching, the points, cited by Blinkstein, about the misunderstanding of what Education and learning is about become particularly relevant.

## 5. Lessons from AI

The key principles of learning outlined in the previous section underpin much of the ambition of the AIED Community's endeavour (see e.g., the International Journal of Artificial Intelligence in Education: https://www.springer.com/journal/40593). Although, in the early days of the field's existence, much effort has been dedicated to finding best ways to automate the didactic model of education (e.g., through drill-and-practice Intelligent Tutoring Systems such as BUGGY – Blando et al., 1989), with the AI system acting as an instructor whose job was to 'diagnose' students for misconceptions and nudge them towards domain expertise, AIED researchers soon realised the limitations of this approach, both pedagogically and technologically. Notably, they started to incorporate into their designs elements of constructionist principles of learning advocated by Papert, and to base their AIED systems on user-centred and/or participatory design methods as a way of aligning them with real-world needs.

This gradual change was both bolstered by the educational research supporting the importance of Papert's work to real educational practices and to learning, and by the realisation that technology underpinned with Wumpus-like algorithms, was too limited to guarantee the accuracy of the data and their interpretation garnered from learner-AI interactions. This realisation has led to some of the most exciting developments not only in AIED, but arguably in human-AI interaction design. This is because by emphasising the importance of human control over the system, they offer early, operational examples of AI transparency and explainability, and of how the *standard model of AI*, whereby AI is in control, can be challenged and replaced with other models, where the user is in charge. In particular, these examples include AIED paradigms such as the *Open Learner Models* (OLMs), where users are given different degrees of control over their data and assessments (Bull & Kay, 2016; Conati, Porayska-Pomsta & Mavrikis, 2018), *teachable agents*, where the users are in charge of what and how is being taught, and *exploratory environments*, where users learn through exploration of the learning domain. Examples of these different paradigms are offered next against the four main lessons that can be derived from the AIED work to date.

One of the overarching lessons learned by the AIED community is the fundamental importance of keeping the human in the loop in the development, evaluation and deployment of the AIED systems, as well as in the way that the interactions are designed. Participation of target users and of relevant stakeholders at different stages in the AI development pipeline has been found to be paramount to the success of the resulting systems, all in terms of their relevance, educational quality and uptake of these systems (Holstein & Doroudi, 2022). The AIED field has been one of the earliest adopters of user-centred and participatory design methods as well as of action research approaches, which has led to critical shifts in the mindset of many AIED researchers from strictly techno-centric to human-centred ones, i.e., ones that are driven by the needs on the ground (Porayska-Pomsta, 2023).

The outcomes of this shift have led quickly to the insights from educational and broader learning sciences domains regarding human cognition, learning and teaching practices providing the backbone of AIED systems. In particular, much work within the field has focused on supporting the development of systems that can support learners in actively constructing their knowledge through tailored activities, interaction and feedback design. Of particular importance are so-called exploratory learning environments (ELEs) such as the ECHOES system (Porayska-Pomsta et al., 2018) or the Fraction Lab (Mavrikis et al., 2022). ELEs allow learners to delve independently into interactive and immersive contexts, to navigate through content, problems or activities and to engage with content in a self-directed, inquisitive and investigative manner. The emphasis of ELEs is on learner autonomy, interactive experiences, and a focus on problem-solving, and on providing opportunities for situated, immersive and engaging experiences, e.g., in simulations or virtual labs or in interactions with socially-endowed embodied agents. This approach prioritises inquiry-based learning, where students formulate and explore their own questions and hypotheses, benefiting from immediate feedback and potentially collaborative learning experiences, thereby nurturing their critical thinking and application of knowledge in varied contexts. The role of AI in these systems is to be able to detect the exploration paths of individual learners and to appropriately adapt its interaction and feedback dynamically and seamlessly.

Teachable agents (e.g., Brophy et al., 1999) present another daring example of an AIED specific approach to fostering active construction of knowledge by learners, wherein students learn by teaching an AI agent. This paradigm, which capitalises on the idea of learning-by-teaching, involves students instructing an agent, guiding it to perform tasks, and helping it to learn. The idea is that by teaching the agent, students are compelled to organise and articulate their knowledge in a coherent and structured manner, which subsequently enhances their retention of information and allows them to transform it into deep understanding and knowledge. The teachable agents' paradigm operates on the premise that teaching something is not only a powerful way to solidify and deepen learning, but also a method to enhance metacognitive skills, as students must reflect on and critically analyse the subject matter to convey it to their artificial pupil effectively. This approach is lauded for its ability to engage students, catalyse deep learning, and foster a constructive learning environment.  In this context, AI facilitates a two-way interactive learning experience, with the intelligent agents also utilising AI to learn from the student, make inferences, and apply acquired knowledge in various problem-solving contexts, all while presenting students with a visible model of their teaching efficacy. Furthermore, the AI enables the agents to exhibit learning, make mistakes, ask clarifying questions, and even help students to reflect on and debug their thought processes.

The metacognitive skills, i.e., skills related to self-awareness and self-regulation promoted both through ELEs and through the teachable agents' paradigms are also the focus of another powerful AIED approach called the Open Learner Models (OLMs).  The OLMs, so-called, because they provide the user with access to the data that they glean about them (their knowledge, skills, and/or emotional and motivational states), have been devised to: (i) enhance the accuracy of the system's inferences about the learner and (ii) allow learners to view and interact with the model as a way of promoting self-understanding and self-regulation.  As such OLMs represent tangible examples of transparent and explainable AI in actual use, with the transparency facilitating deep, tailored learning experiences, enabling educators to customise their support, and empower students to engage actively in their learning by inspecting the AI's assessment of their learning, by negotiating the AI assessment of them, and by even challenging the system's perception of their knowledge and abilities (Bull & Kay, 2016). Consequently, OLMs act as a multifaceted tool in educational technology, aiding in personalising education while simultaneously bolstering learners' self-awareness and metacognitive development (*ibid*; Porayska-Pomsta & Chryssafidou, 2018).

In this context AI is used to manage and adapt algorithmically the representation of a learner's knowledge, skills, and other attributes. AI algorithms analyse the learner's interactions, responses, and performance data, to update the OLM dynamically. This updated model subsequently informs personalised learning experiences, ensuring that the content, feedback, and challenges presented to the learner are tailored to their current understanding and capabilities. AI also facilitates the OLM-user interactions, e.g., through visualisations to make learner data interpretable and interactive for users (learners, teachers, or peers), which can subsequently influence the selection of the learning strategies, self-regulation, and instructional interventions within an AIED, as well as traditional learning contexts.

One of the challenges in supporting active engagement in students' own learning, their metacognitive skills and self-regulation, lies in the way that the relevant pedagogic support is tailored to their moment-by-moment needs. Adaptation of feedback in terms of content and the timing of the delivery constitutes a critical part of such support.  Finding the appropriate feedback strategies and tactics that are able to prop and guide learners' active construction of knowledge constitutes a central part of the AIED research. A critical aspect of this research concerns the idea of learners' help-seeking behaviours as a way of supporting their productive engagement and rehearsal of metacognitive competencies. Research in this domain often focuses on creating systems that foster both independent learning and effective help-seeking when needed and on discerning between strategic help-seeking (a legitimate, learning-oriented strategy) and inappropriate or misuse of help, such as help abuse (e.g., gaming the system to get to the answer fast). Accordingly, scholars and developers have explored adaptive help-seeking frameworks that consider not only the learner's knowledge, skills and their metacognitive and motivational states. In particular, systems such as the Cognitive Tutors[7] employ models that track and analyse students' behaviours and performance to discern when to offer help, what type of help to provide (e.g., hints, explanations, or redirecting to prior relevant material), and how to present this help in a way that maintains the learner's engagement and motivation, while avoiding the purely didactic approach to pedagogical support. In this case AI serves to predict when students might need help, based on historical data and observed behaviour patterns. Others may utilise real-time analytics to assess and respond to learners' immediate needs. The ultimate objective across these approaches is to cultivate effective help-seeking strategies among learners, ensuring that they can navigate challenges productively and develop and apply new knowledge and skills.

## 6. Discussion and Conclusions

Exploring fifty years of the AIED research reveals a compelling narrative of opportunities and challenges as it combines technological potential with educational research and praxis. AIED has pioneered in challenging both the established AI models and their implementations, and educational methods deemed inadequate to provide long-lasting foundations for deep and life-long learning or the motivation to learn. Its research offers tangible examples, backed with efficacy evaluations of how learners can be put in a position of productive control over their own learning – a position which has been demonstrated to produce long lasting positive effects and highly desirable outcomes of self-efficacy, curiosity, creativity and a passion for learning. Simultaneously, AIED research also exposes pertinent ethical, philosophical, and practical dilemmas regarding educational fundamentals, human interaction, and learner autonomy. The enthusiasm and caution enveloping AI application in education must be prudently managed by reference to what we know and through inclusion of key stakeholders such as educators and learners themselves, to ensure that we chart a future where educational technology is not only advanced, but also ethically and pedagogically sound.

Addressing the complex challenges at the confluence of AI and education warrants a nuanced approach that synergistically integrates technological, educational, and ethical perspectives – a synergy which is so far lacking. The current notable misalignment between AIED research and practical AI technology deployment in educational settings, and discourse around the AIED risks to misdirect the decisions in this domain. Pivotal to avoiding such misdirection is the need to recognise the rich diversity of AI systems and their impacts, and to be wary of oversimplifying of 'Education' as a system and as a process, since the personal and collective educational experiences and dynamics are not universally applicable or understood. To engender concrete and informed action, a context-specific understanding of educational dynamics is essential, as is an understanding of multiple conflicting stakeholder incentives within the educational ecosystem.

Navigating 50 years in AI in Education (AIED) shows a rich history but also brings to light significant challenges and lessons that are crucial for shaping future learning. The key takeaway from this journey is the critical importance of balancing advanced tech with genuine, human-centred learning experiences. The technocentric viewpoint, particularly one that is steeped in didactic methods and reinforcement techniques, may lead us to being entrenched in an educational paradigm that emphasises rote learning over deep understanding and a passion for lifelong learning. The urgency for informed guidance from education experts and for corresponding governance in AIED cannot be overstated. Despite the burgeoning public and policy discourse on AI governance, a tangible, informed, and anticipatory approach to regulating AIED, that is rooted in a deep understanding of its practical and theoretical aspects, remains conspicuously absent, thereby perpetually relegating us to a reactive instead of pro-active stance in the face of the developing AIED landscapes.

---

[7] https://www.carnegielearning.com/pages/whitepaper-report/the-cognitive-tutor-applying-cognitive-science-to-education/

In looking forward towards how learning of the next generations of learners might be enriched rather than hindered by AI, invites us to drive the technological advancements and the way we deploy technologies in education (if indeed we deem this essential in all circumstances) by a grounded understanding of real-world education and the needs at the front-line of learning and teaching. Our shared knowledge and experiences in AIED provide essential tools for steering us towards a future where technology enhances and genuinely adapts to (rather than dictates) what, when and how we as individuals learn.

## 7. References

Blando, J. A., Kelly, A E., Schneider, Beth R., and Sleeman, D. (1989). Analyzing and Modeling Arithmetic Errors, Journal for Research in Mathematics Education, Vol. 20, No. 3 (May, 1989), pp. 301-308

Blinkstein, P. (2013). Seymour Papert's legacy: Thinking about learning, and learning about thinking, Seymour Papert Tribute at IDC 2013

Bostrom, N (2017). Superintelligence, Oxford University Press.

Brophy, S., Biswas, G., Katzleberger, T., Bransford, J., Schwartz, D. (1999). Teachable Agents: Combining Insights from Learning Theory and Computer Science, https://api.semanticscholar.org/CorpusID:1100669

Bull, S., & Kay, J. (2016). SMILI☺: a framework for interfaces to learning data in open learner models, learning analytics and related felds. International Journal of Artifcial Intelligence in Education,26(1), 293–331.

Clinton, H. (2023). The Weaponization of Loneliness, in the Atlantic, https://www.theatlantic.com/ideas/archive/2023/08/hillary-clinton-essay-loneliness-epidemic/674921/ (accessed 8th of October, 2023).

Conati, C., Porayska-Pomsta, K., & Mavrikis, M. (2018). AI in Education needs interpretable machine learning: Lessons from Open Learner Modelling. ArXiv:1807.00154 [Cs]. http://arxiv.org/abs/1807.00154

Crawford K. (2017). Hidden Biases in Big Data, in Analytics and Data Science Harvard Business Review, https://hbr.org/2013/04/the-hidden-biases-in-big-data (accessed 1 September 2021).

Dubnick, M J. (2014). Toward an Ethical Theory of Accountable Governance. International Political Science Association meeting, July 19-24, Montreal

Gentile, M., Città, G., Perna, S., Allegra, M. (2023). Do we still need teachers? Navigating the paradigm shift of teachers' role in the AI era, Front. Educ., 31 March 2023, Sec. Digital Learning Innovations Vol. 8, https://doi.org/10.3389/feduc.2023.1161777

Holstein, K., & Doroudi, S. (2022). TITLE In W. Holmes, & K. Porayska-Pomsta (Eds.), The ethics of artifcial intelligence in education: practices, challenges, and debates. Routledge

Jobin, A., Ienca, M., & Vayena, E. (2019). Artificial Intelligence: The global landscape of ethics guidelines. Nature Machine Intelligence, 1(9), 389–399. https://doi.org/10.1038/s42256-019-0088-2Kant, 1785.

Jumper, J., Evans, R., Pritzel, A. *et al.* Highly accurate protein structure prediction with AlphaFold. *Nature* **596**, 583–589 (2021). https://doi.org/10.1038/s41586-021-03819-2

Mavrikis, M., Rummel, N., Wiedmann, M. *et al.* (2022). Combining exploratory learning with structured practice educational technologies to foster both conceptual and procedural fractions knowledge. *Education Tech Research Dev* **70**, 691–712. https://doi.org/10.1007/s11423-022-10104-0

Mazzucato, M., Schaake, M., Krier, S., & Entsminger, J. (2022). Governing artifcial intelligence in the public interest. UCL institute for innovation and public purpose, working paper series (IIPP WP 2022-12). https://www.ucl.ac.uk/bartlett/public-purpose/wp2022-12

Nti Asare, I. (2023). Working Together: *Policy Design for Collective Action in AIED*. Unpublished PhD Upgrade Chapter, University College London, 2023

Papert, S. (1994). The Children's Machine: Rethinking School in the Age of the Computer, Basic Books, 978-0465010639.

Pedro, F., Subosa, M., Rivas, A., and Valverde, P. (2019). Artificial intelligence in education: Challenges and opportunities for sustainable development. Technical report.

Porayska-Pomsta, K. (2016) AI as a Methodology for Supporting Educational Praxis and Teacher Metacognition. *International Journal of Artificial Intelligence in Education,* **26**, 679–700 (2016). https://doi.org/10.1007/s40593-016-0101-4

Porayska-Pomsta, K. et al. (2018). Blending human and artificial intelligence to support children's social communication skills, ACM Transactions on Computer-Human Interaction, Vol. 25, No. 6, Article 35, https://doi.org/10.1145/3271484

Porayska-Pomsta, K. and Chrissafidou, E. (2018). Adolescents' self-regulation during job interviews through an AI coaching environment, Artificial Intelligence in Education: 19th International Conference, AIED 2018, London, UK, June 27–30, 2018, Proceedings, Part II 19

Porayska-Pomsta, K. and Rajendran, T. (2019). Accountability in human and artificial intelligence decision-making as the basis for diversity and educational inclusion, Artificial Intelligence and Inclusive Education: Speculative Futures and Emerging Practices, 39-59, Springer Singapore

Porayska-Pomsta, K., Holmes, W. and Nemorin, S. (2023). Handbook of Artificial Intelligence in Education

Porayska-Pomsta (2023). *A manifesto for a pro-actively responsible AIED*, International Journal of Artificial Intelligence in Education, Holmes and Kay (eds.), Springer Nature.

Russell, S. (2019). Human compatible: artificial intelligence and the problem of control, Penguin, ISBN:9780525558620.

Russell, S. and Norvig, P. (2020). Artificial Intelligence: A Modern Approach, 4th Global Edition, Pearson, ISBN 978-0134610993

UNESCO (2021). AI and education: guidance for policy-makers, https://unesdoc.unesco.org/ark:/48223/pf0000376709

UNESCO (2022). Recommendations on Ethics of Artificial Intelligence, https://www.unesco.org/en/articles/recommendation-ethics-artificial-intelligence

UNESCO (2023). The teachers we need for the education we want: the global imperative to reverse the teacher shortage; factsheet, code ED/PLS/TED/2023/03

UNESCO (2023b). Guidance for Generative AI in Education and research. https://unesdoc.unesco.org/ark:/48223/pf0000386693

Woolf, B, (2008). Building Intelligent Tutoring Systems. Morgan Kaufman.